\documentclass{iau}
\usepackage{graphicx}

\title[M87 radio and VHE connection in 2012] 
{A strong radio brightening at the jet base of M87 during the elevated
very-high-energy $\gamma$-ray state in 2012}

\author[K. Hada, M. Giroletti, M. Kino et al.]   
{
K.~Hada$^{1,2}$
M.~Giroletti$^1$, 
M.~Kino$^{3,4}$, 
G.~Giovannini$^{1,5}$, 
F.~D' Ammando$^1$,
C.~C.~Cheung$^6$,
M.~Beilicke$^7$, 
H.~Nagai$^7$, 
A.~Doi$^4$,
K.~Akiyama$^{2,9}$, 
M.~Honma$^{2,10}$,
K.~Niinuma$^{10}$, 
C.~Casadio$^{12}$,
M.~Orienti$^1$, 
H.~Krawczynski$^7$, 
J.~L.~G\'omez$^{12}$, 
S.~Sawada-Satoh$^2$, 
S.~Koyama$^{2,4,9}$, 
A.~Cesarini$^{13}$, 
S.~Nakahara$^{14}$ 
 \and 
M.~A.~Gurwell$^{15}$
}

\affiliation{
$^1$INAF Istituto di Radioastronomia, via Gobetti 101, I-40129 Bologna, Italy \\
email: {\tt kazuhiro.hada@nao.ac.jp} \\
[\affilskip]
$^2$Mizusawa VLBI Observatory, National Astronomical Observatory of Japan,
Osawa, Mitaka, Tokyo 181-8588, Japan \\
$^3$Korea Astronomy and Space Science Institute, 776 Daedukdae-ro, Yusong,
Daejon 305-348, Korea \\
$^4$Institute of Space and Astronautical Science, Japan Aerospace
Exploration Agency, 3-1-1 Yoshinodai, Chuo, Sagamihara 252-5210, Japan \\
$^5$Dipartimento di Fisica e Astronomia, Universit\`a di Bologna, via
Ranzani 1, I-40127 Bologna, Italy \\
$^6$Space Science Division, Naval Research Laboratory, Washington, DC
20375, USA \\
$^7$Physics Department and McDonnell Center for the Space Sciences,
Washington University, St. Louis, MO 63130, USA \\
$^8$National Astronomical Observatory of Japan, Osawa, Mitaka, Tokyo
181-8588, Japan \\
$^9$Department of Astronomy, Graduate School of Science, The University of
Tokyo, 7-3-1 Hongo, Bunkyo-ku, Tokyo 113-0033, Japan \\
$^{10}$Department of Astronomical Science, The Graduate University for
Advanced Studies (SOKENDAI), 2-21-1 Osawa, Mitaka, Tokyo 181-8588, Japan \\
$^{11}$Graduate School of Science and Engineering, Yamaguchi University, 1677-1
Yoshida, Yamaguchi, 753-8512, Japan \\
$^{12}$Instituto de Astrofisica de Andalucia, CSIC, Apartado 3004, 18080
Granada, Spain \\
$^{13}$Department of Physics, University of Trento, I38050, Povo, Trento,
Italy \\
$^{14}$Faculty of Science, Kagoshima University, 1-21-35 Korimoto, Kagoshima,
Kagoshima 890-0065, Japan \\
$^{15}$Harvard-Smithsonian Center for Astrophysics, Cambridge, MA 02138, USA
}

\pubyear{2014}
\volume{313} 
\pagerange{xx--xx}
\jname{Extragalactic jets from every angle}
\editors{F. Massaro, C.C. Cheung, E. Lopez, A. Siemiginowska, eds.}

\begin{document}

\maketitle

\begin{abstract}
The nearby radio galaxy M87 offers a unique opportunity for exploring the
connection between $\gamma$-ray production and jet formation at an unprecedented
linear resolution. However, the origin and location of the $\gamma$-rays in this
source is still elusive. Based on previous radio/TeV correlation events, the
unresolved jet base (radio core) and the peculiar knot HST-1 at $>$120~pc from the
nucleus are proposed as candidate site(s) of $\gamma$-ray production. Here we report
our intensive, high-resolution radio monitoring observations of the M87 jet with
the VLBI Exploration of Radio Astrometry (VERA) and the European VLBI Network
(EVN) from February 2011 to October 2012. During this period, an elevated level of
the M87 flux is reported at TeV with VERITAS. We detected a remarkable flux
increase in the radio core with VERA at 22/43 GHz coincident with the VHE
activity. Meanwhile, HST-1 remained quiescent in terms of its flux density and
structure at radio. These results strongly suggest that the TeV $\gamma$-ray activity
in 2012 originates in the jet base within 0.03~pc (projected) from the central
supermassive black hole.  
\end{abstract}

\firstsection 
\section{Introduction}
The nearby radio galaxy M~87 accompanies one of the best studied AGN jets. Its
proximity (16.7~Mpc) and brightness have enabled intensive studies of this jet
over decades from radio, optical and to X-ray at tens of parsec scale
resolutions. Furthermore, the inferred very massive black hole~($M_{\rm BH} \simeq
(3-6) \times 10^9~M_{\odot}$) yields a linear resolution down to 
$1~{\rm milliarcsecond~(mas)}=0.08~{\rm pc}=140$ Schwarzschild
radii~$(R_{\rm s})$ (for $M_{\rm BH} = 6 \times 10^9~M_{\odot}$), 
making this source an ideal case to probe the
relativistic-jet formation at an unprecedented compact scale with
Very-Long-Baseline-Interferometer (VLBI) observations~(e.g., Hada et al. 2011;  
Asada \& Nakamura 2012; Doeleman et al. 2012; Hada et al. 2013).

M~87 is now widely known to show $\gamma$-ray emission up to the very-high-energy
(VHE; $E>100$~GeV) regime, where this source often exhibits active flaring
episodes. The location and the physical processes of such emission have been a
matter of debate over the past years, and there are two candidate sites which can
be responsible for the VHE $\gamma$-ray production. One is a very active knot
HST-1 which is located at more than 100 pc from the
nucleus~(Stawarz et al. 2006; Cheung et al. 2007; Harris et al. 2009). This
argument is based on the famous VHE flare event in 2005, where HST-1 underwent the
large radio-to-X-ray outburst jointly with a VHE flare. In contrast, the other
candidate is the core/jet base, which is very close to the central black
hole. This argument is based on the TeV event in 2008, where the core/TeV showed a
remarkable correlation in the light curves. There was another VHE event in 2010, but this is rather
elusive. Coincident with the VHE event, \textit{Chandra} detected an
enhanced flux from the X-ray core~(Harris et al. 2011; Abramowski et al. 2012), and VLBA
observations also suggested a possible increase of the radio core
flux~(Hada et al. 2012). However, Giroletti et al. (2012) found the emergence of a
superluminal component in the HST-1 complex near the epoch of this event, which is
reminiscent of the 2005 case.

Recently, the VERITAS Collaboration has reported new VHE $\gamma$-ray activity
from M~87 in early 2012~(Beilicke et al. 2012). While there were no remarkable flares
like those in the previous episodes, the VHE flux in 2012 clearly
exhibits an elevated state at a level of $\sim$$9\sigma$ ($\Phi_{\rm >
0.35TeV}\sim$(0.2--0.3)$\times 10^{-11}$ photons~cm$^{-2}$~s$^{-1}$) over the
consecutive two months from February to March 2012. The observed flux is a factor
of $\sim$2 brighter than that in the neighboring quiescent periods. 
Therefore, this event provides another good opportunity for exploring the location of the 
VHE emission site by jointly using high-resolution instruments. 

\section{Observations}
Here we report a multi-wavelength radio and MeV/GeV study of the M~87 jet during
this period using the VLBI Exploration Radio Astrometry (VERA), the European VLBI
Network (EVN), the Submillimeter Array (SMA) and the \textit{Fermi}-LAT. We
especially focus on the VLBI data in the radio bands; with VERA, we obtained the
high-angular-resolution, dense-sampling-interval, phase-referencing data set at 22
and 43~GHz during the VHE activity in 2012: with the supportive EVN monitoring, we
obtained a complementary data set at 5~GHz, which enables a high-sensitivity
imaging of the M~87 jet. A collective set of these radio data allows us to probe
the detailed physical status and structural evolutions of M~87 by pinpointing the
candidate sites of the $\gamma$-ray emission i.e., the core and HST-1. For more
details regarding the data analysis, see Hada et al.~(2014).

\begin{figure}[htbp]
\begin{center}
 \scalebox{1}{\includegraphics{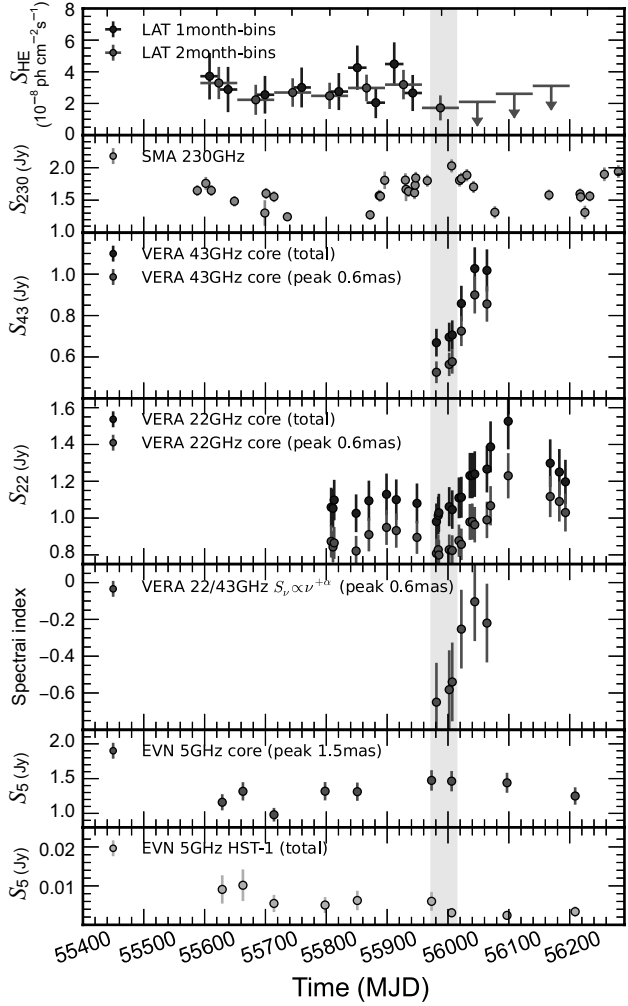}} 
 \caption{Multi-wavelength light curves of M~87 between February 2011 and December 2012. The vertical shaded area over the plots indicates a period of elevated VHE emission reported by \cite{beilicke2012}.} \label{fig:}
\end{center}
\end{figure}

\section{Results}
In Figure~1 we show a combined set of light curves of M~87 from radio to MeV/GeV
$\gamma$-ray between MJD~55400 and MJD~56280. Thanks to the dense, complementary
coverages of VERA and EVN, we revealed the detailed evolutions of the radio light
curves for both the core and HST-1. The most remarkable finding in these plots is
a strong enhancement of the radio core flux at VERA 22 and 43~GHz, which occurred
coincidentally with the elevated VHE state. At 22~GHz, we further detected a
subsequent decay stage of the brightness at the last three epochs. Also at 43~GHz,
we detected possible saturation of the flux increase near the last
epoch. Meanwhile, the EVN monitoring confirmed a constant decrease of the HST-1
luminosity.  Figure 2 describes VERA 43-GHz images during the VHE active period, which indicate 
the flux enhancement within the central resolution element of 0.4~mas, corresponding to a linear scale
of 0.03 pc or 56~$R_{\rm s}$. We also note that the SMA data at 230~GHz also appear to show a local maximum in its light curve during the period of the elevated VHE state. 

Another notable finding is a frequency-dependent evolution of
the radio core flare. The VERA light curves clearly indicate that the radio
core brightens more rapidly with a larger amplitude as frequency increases. At
43~GHz, the flux increased up to $\sim$70\% for the subsequent 2 months 
at an averaged rate of $\sim$35\%/month, and afterward
the growth seems to be saturated. On the other hand, the core flux at 22-GHz
progressively increased up to $\sim$50\% for the subsequent 4 months 
at a slower rate of $\sim$12\%/month. At 5~GHz, by contrast, the core
remained virtually stable within the adopted error of 10\%. This is the first time
that such a frequency-dependent nature of the radio flare is clearly confirmed in
the M~87 jet. We also detected a core-shift between 22 and 43GHz by using the VERA dual-beam astrometry technique (see Hada et al. 2014), 
where the shift amount was similar to the value obtained in the previous core-shift 
measurement (Hada et al. 2011). 

Regarding the MeV/GeV regime, the LAT light curves were stable up to February 2012, 
and we did not find any significant flux enhancement during the period of the VHE activity.  
After March 2012, however, no significant emission
was detected for the subsequent 6 months in the 1- and 2-month binned data,
suggesting a change in the HE state after the VHE event. 
This indicates a decrease in the HE flux (by a factor of $\sim
2$) after the VHE event, in agreement with the level of decrease observed at VHE in
2012 April-May (Beilicke et al. 2012).

\begin{figure}[t]
\begin{center}
 \scalebox{0.3}{\includegraphics{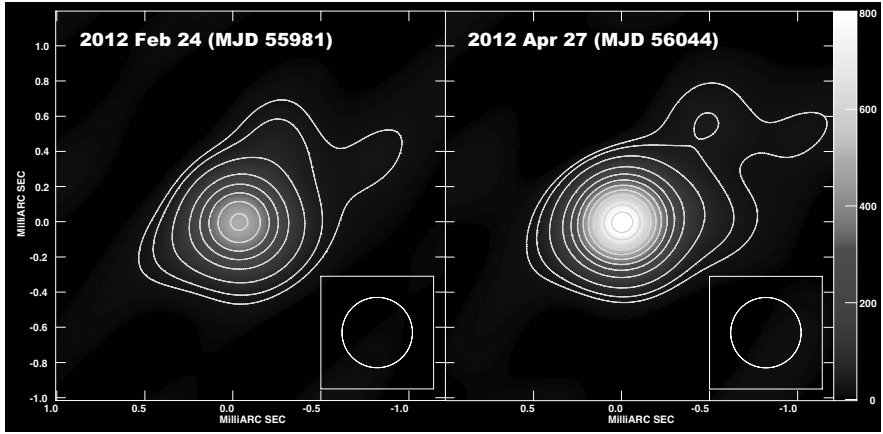}} 
 \caption{VERA 43 GHz images of the M87 jet during the elevated VHE state in 2012.} \label{fig:}
\end{center}
\end{figure}

\section{Discussion and Summary}
Following the 2008 episode this is the second time where a VHE event
accompanied a remarkable radio flare from the core. Meanwhile, the radio
luminosity of the HST-1 region was continuously decreasing, and we did not find
any hints of the emergence of new components from HST-1 as seen in 2005 and
2010. These results strongly suggest that the VHE activity in 2012 is associated
with the core at the jet base, while HST-1 is an unlikely source. We note
that these remarkable flares are very rare also in radio bands~(Acciari et al. 2009), 
so it is unlikely that an observed joint
radio/VHE correlation is a chance coincidence, while the low statistics of the LAT
light curves still do not allow conclusive results on the HE/VHE
connection. 

What kinds of mechanisms are responsible for the VHE production in the M87 core? 
This question is still elusive. 
Some of the existing models ascribe the VHE production to extremely compact
regions near the central black hole (e.g., Neronov \& Aharonian 2007; Lenain et al. 2008; 
Giannios et al. 2010; Barkov et al. 2010). 
These models well explain the rapid (a few days) variability
observed in the previous VHE flares in 2005, 2008 and 2010.  
However as far as we consider the case in 2012, the size of the
associated region expected from these models seems to be
smaller than that suggested by VLBI and the observed longer
timescale of the VHE variability. Indeed, a contemporaneous mm-VLBI observation
at 230~GHz during the 2012 event also suggests the possible extended nature  
for the flaring region ($\gtrsim$0.3~mas; Akiyama et al. in prep.).

Another popular scenario for the M87 VHE production comes from a blazar-type, 
two-zone emission model where 
the VHE emission originates in the upstream part of a decelerating jet (Georganopoulos et al. 2005) or in the layer part of the spine-sheath structure (Tavecchio \& Ghisellini 2008). 
However in their steady state models, whether the models can
explain the observed simultaneous radio/VHE correlation or not has not been well
investigated yet because the emission regions associated with radio and VHE are
spatially separated from each other. In this respect, a simple,
homogeneous one-zone synchrotron self-Compton jet model examined by Abdo et al. (2009) 
would be interesting to note since one can in principle accept coincident radio/VHE correlations in this
context. 

Our multi-frequency radio monitoring additionally revealed a frequency-dependent evolution of
the radio light curves for the M~87 core. Such a behavior is often explained by the creation
of a plasma condensation,
which subsequently expands and propagates down the jet under the effect of
synchrotron self-absorption (SSA). The stronger SSA opacity at
the jet base causes a delayed brightening at lower frequencies, and the light curve at each frequency 
reaches its maximum when the
newborn component passes through the $\tau_{\rm ssa}(\nu)\sim 1$ surface (i.e.,
the radio core at the corresponding frequency). In this context, by jointly using the observed time-lag ($\Delta t_{\rm 43-22}$) and core-shift ($\Delta r_{\rm proj, 43-22}$), 
we can estimate an apparent speed of the propagating component 
as $\beta_{\rm app, 43 \rightarrow 22} = \frac{\Delta r_{\rm proj, 43-22}}{c \Delta
 t_{\rm 43-22}}$. This results in a speed about $\sim$0.04$c$--0.22$c$, 
suggesting that the newborn component is sub-relativistic.   
This is significantly smaller than the super-luminal features appeared from the core
during the previous VHE event in 2008~(1.1$c$; Acciari et al. 2009), where
the peak VHE flux is $\gtrsim$5 times higher than that in 2012. If we assume that
propagating shocks or component motions seen in radio observations reflect the
bulk velocity flow, this may suggest that the stronger VHE activity is associated
with the production of the higher Lorentz factor jet.

We are currently upgrading our 
M87 monitoring project by using the KVN and VERA Array (KaVA; Niinuma et al. 2014), which 
dramatically improves jet imaging capability thanks to the increase of the number of telescopes/baselines, 
and the inclusion of shorter baselines. This will allow us to constrain the jet kinematics and radio/VHE connection more precisely.

\section*{Acknowledgments}
The VERA is operated by Mizusawa VLBI Observatory, a branch of National
Astronomical Observatory of Japan. The Submillimeter Array is a joint project
between the Smithsonian Astrophysical Observatory and the Academia Sinica
Institute of Astronomy and Astrophysics and is funded by the Smithsonian
Institution and the Academia Sinica.

The \textit{Fermi} LAT Collaboration acknowledges generous ongoing support from a
number of agencies and institutes that have supported both the development and the
operation of the LAT as well as scientific data analysis. These include the
National Aeronautics and Space Administration and the Department of Energy in the
United States, the Commissariat \`a l'Energie Atomique and the Centre National de
la Recherche Scientifique Institut National de Physique Nucl\'eaire et de Physique
des Particules in France, the Agenzia Spaziale Italiana and the Istituto Nazionale
di Fisica Nucleare in Italy, the Ministry of Education, Culture, Sports, Science
and Technology (MEXT), High Energy Accelerator Research Organization (KEK) and
Japan Aerospace Exploration Agency (JAXA) in Japan, and the K.~A.~Wallenberg
Foundation, the Swedish Research Council and the Swedish National Space Board in
Sweden.

Additional support for science analysis during the operations phase is gratefully
acknowledged from the Istituto Nazionale di Astrofisica in Italy and the Centre
National d'\'Etudes Spatiales in France.

\end{document}